\documentclass[9pt,conference]{IEEEtran}
\IEEEoverridecommandlockouts

\usepackage{cite}
\usepackage{multirow}
\usepackage{amsmath,amssymb,amsfonts,subcaption,hyperref}
\usepackage{algorithmic}
\usepackage{graphicx}
\usepackage{textcomp}
\usepackage{xcolor}
\def\BibTeX{{\rm B\kern-.05em{\sc i\kern-.025em b}\kern-.08em
    T\kern-.1667em\lower.7ex\hbox{E}\kern-.125emX}}

\DeclareMathOperator*{\argmin}{arg\,min}

\begin{document}

\title{Adversarial Deep-Unfolding Network for MA-XRF Super-Resolution on Old Master Paintings Using Minimal Training Data\\

\thanks{$^{*}$Herman Verinaz-Jadan and Su Yan contributed equally to this work.}
\thanks{This work is in part supported by EPSRC grant EP/R032785/1.}
}

\author{\IEEEauthorblockN{Herman Verinaz-Jadan\IEEEauthorrefmark{2}$^{*}$, Su Yan\IEEEauthorrefmark{3}$^{*}$, Catherine Higgitt\IEEEauthorrefmark{4}, and Pier Luigi Dragotti\IEEEauthorrefmark{5}}
\IEEEauthorblockA{\IEEEauthorrefmark{2}Faculty of Electrical and Computer Engineering (FIEC)\\ Escuela Superior Politecnica del Litoral (ESPOL), Ecuador\\
Email: hverinaz@espol.edu.ec}
\IEEEauthorblockA{\IEEEauthorrefmark{3}Department of Bioengineering,	Imperial College London, UK\\
Email: s.yan18@imperial.ac.uk}
\IEEEauthorblockA{\IEEEauthorrefmark{4}Scientific Department, The National Gallery, London, UK\\
Email: catherine.higgitt@nationalgallery.org.uk}
\IEEEauthorblockA{\IEEEauthorrefmark{5}Department of Electrical and Electronic Engineering, Imperial College London, UK\\
Email: p.dragotti@imperial.ac.uk}
}

\maketitle

\begin{abstract}
High-quality element distribution maps enable precise analysis of the material composition and condition of Old Master paintings. These maps are typically produced from data acquired through Macro X-ray fluorescence (MA-XRF) scanning, a non-invasive technique that collects spectral information. However, MA-XRF is often limited by a trade-off between acquisition time and resolution. Achieving higher resolution requires longer scanning times, which can be impractical for detailed analysis of large artworks. Super-resolution MA-XRF provides an alternative solution by enhancing the quality of MA-XRF scans while reducing the need for extended scanning sessions. This paper introduces a tailored super-resolution approach to improve MA-XRF analysis of Old Master paintings. Our method proposes a novel adversarial neural network architecture for MA-XRF, inspired by the Learned Iterative Shrinkage-Thresholding Algorithm. It is specifically designed to work in an unsupervised manner, making efficient use of the limited available data. This design avoids the need for extensive datasets or pre-trained networks, allowing it to be trained using just a single high-resolution RGB image alongside low-resolution MA-XRF data. Numerical results demonstrate that our method outperforms existing state-of-the-art super-resolution techniques for MA-XRF scans of Old Master paintings.
\end{abstract}


\section{Introduction}
\label{sec:intro}

Macro X-ray fluorescence (MA-XRF) scanning is a method used in the analysis of materials and artistic techniques in Old Master paintings, revealing the distribution of chemical elements across the artwork. Specifically, MA-XRF measures secondary X-ray photons emitted from elements when illuminated by a primary X-ray beam~\cite{NG_Technical_41,yan2020revealing,yan2021prony}. This process produces a data array that captures both the spatial and spectral information of the painting~\cite{10096322}.

To achieve high-quality MA-XRF data with high signal-to-noise ratio (SNR) and spatial resolution, traditional methods extend the scan time per spot and minimize the X-ray beam diameter. While effective, these adjustments significantly increase the overall scanning duration. In contrast, MA-XRF super-resolution (SR) techniques enhance high-resolution (HR) MA-XRF images from low-resolution (LR) data post-collection, reducing the need for prolonged data collection and thereby simplifying the process~\cite{dai2017spatial,10096322}.

In MA-XRF image super-resolution (SR), initial work primarily focused on non-deep learning methods. Dai et al. propose dictionary learning to integrate HR RGB with LR MA-XRF images \cite{dai2017spatial}. This approach aligns with broader efforts in hyperspectral imaging (HSI) super-resolution, where techniques such as sparse representation, matrix factorization, and tensor factorization have been used to enhance HR multispectral imaging (MSI) data or HR RGB images \cite{li2018fusing,zhang2018exploiting,kanatsoulis2018hyperspectral,dian2019learning,wan2020nonnegative,xue2021spatial,he2022hyperspectral}. Subsequently, a method applying coupled dictionary learning was proposed in \cite{10096322}, effectively distinguishing and exploiting the shared and unique information in HR RGB and MA-XRF images to prevent the incorporation of artificial features.

The introduction of deep learning has expanded the approaches available for addressing super-resolution challenges in spectral imaging, including HSI. Several methods based on deep neural networks (DNNs) have been proposed \cite{zhang2020unsupervised,wei2020deep,zhu2020hyperspectral,wang2021hyperspectral,hu2021hyperspectral,dong2021model,liu2022model} to solve HSI super-resolution problems. However, the direct application of DNN techniques to Macro X-ray Fluorescence presents challenges, primarily due to differences in spectral ranges and lack of training data. These discrepancies highlight the need for developing deep-learning approaches specifically tailored to MA-XRF data.

In this work, we propose a model-based deep learning approach specifically designed for the MA-XRF super-resolution problem. Our network, inspired by the Learned Iterative Shrinkage-Thresholding Algorithm (LISTA)~\cite{gregor2010learning}, effectively exploits HR information from RGB images to improve LR MA-XRF scans. It does not depend on additional training data or pre-trained models, which simplifies its implementation and enhances its practicality for art analysis.

\section{Problem Formulation}

This problem requires to reconstruct a HR MA-XRF image by leveraging information from both MA-XRF and RGB modalities. Specifically, the reconstruction process involves synthesizing an HR MA-XRF image, $\mathbf{Y}~\in~\mathbb{R}^{B \times N_{\text{h}}}$, from a given LR MA-XRF image, $\mathbf{Y}_{\downarrow}~\in~\mathbb{R}^{B \times N_{\text{l}}}$, and an HR RGB image, $\mathbf{Z} \in \mathbb{R}^{b \times N_{\text{h}}}$. Here, $B$ represents the number of spectral channels in the MA-XRF data, $N_{\text{h}} = H \times W$ and $N_{\text{l}} = h \times w$ denote the total number of pixels of the HR and LR images, respectively, where $H$ and $W$ are the height and width of the HR images, and $h$ and $w$ are the height and width of the LR images.

To facilitate the reconstruction, a dictionary-based representation is commonly used to capture the intrinsic characteristics shared between the MA-XRF and RGB images as well as their unique features~\cite{6200362,974727,10.1117/12.331889,4011956,1710377, 10096322}. In this work, we formulate this problem within the context of a multimodal super-resolution framework, following the conventions in~\cite{10096322}. Thus, the HR MA-XRF and RGB images are represented with a set of dictionaries, capturing both shared (common) and distinct (unique) features, as follows:
    \begin{align}\label{eq:problemF}
    \begin{cases}
        \mathbf{Y}=\mathbf{Y}_{\text{c}} + \mathbf{Y}_{\text{u}}=\mathbf{D}_{\text{c}}^{\text{xrf}}\mathbf{A}_{\text{c}}+\mathbf{D}_{\text{u}}^{\text{xrf}}\mathbf{A}_{\text{u}}^{\text{xrf}}, \\
        \mathbf{Z}=\mathbf{Z}_{\text{c}} + \mathbf{Z}_{\text{u}}=\mathbf{D}_{\text{c}}^{\text{rgb}}\mathbf{A}_{\text{c}}+\mathbf{D}_{\text{u}}^{\text{rgb}}\mathbf{A}^{\text{rgb}}_{\text{u}},
    \end{cases}
    \end{align}
where $\mathbf{Y}$ represents the HR MA-XRF image and $\mathbf{Z}$ the HR RGB image. The subscript $\text{c}$ denotes the common components shared between the MA-XRF and RGB modalities, while $\text{u}$ denotes the unique components specific to each modality. $\mathbf{D}_{\text{c}}^{\text{xrf}}  \in \mathbb{R}^{B \times M_{c}}$ and $\mathbf{D}_{\text{u}}^{\text{xrf}} \in \mathbb{R}^{B \times M_{xu}}$ are the dictionaries containing $M_{c}$ and $M_{xu}$ atoms, respectively. Similarly, $\mathbf{D}_{\text{c}}^{\text{rgb}}  \in \mathbb{R}^{b \times M_{c}}$ and $\mathbf{D}_{\text{u}}^{\text{rgb}}   \in \mathbb{R}^{b \times M_{ru}}$ serve the same purpose for the RGB data. The matrices $\mathbf{A}_{\text{c}}$, $\mathbf{A}_{\text{u}}^{\text{xrf}}$, and $\mathbf{A}^{\text{rgb}}_{\text{u}}$ are the representation matrices for the common and unique components, with $\mathbf{A}_{\text{c}}$ shared across both modalities.

Furthermore, the LR MA-XRF image, $\mathbf{Y}_{\downarrow}$, is acquired from the HR MA-XRF image, $\mathbf{Y}$, by applying the downsampling matrix $\mathbf{U}$:
    \begin{equation}\label{eq:problemF2}
        \mathbf{Y}_{\downarrow}
=\mathbf{Y}\mathbf{U}.
    \end{equation}
This formulation facilitates the application of dictionary learning techniques for super-resolution~\cite{dai2017spatial, 10096322}. The problem is therefore to reconstruct the HR MA-XRF image, $\mathbf{Y}$, given the relationships defined in Equations~\eqref{eq:problemF} and~\eqref{eq:problemF2}, along with the available LR MA-XRF image, $\mathbf{Y}_{\downarrow}$, and HR RGB image, $\mathbf{Z}$.

Note that we can consolidate Equations in~\ref{eq:problemF} into a single representation, $\mathbf{A}$, employing a concatenated vector $\mathbf{X}$ and an expanded dictionary $\mathbf{D}$:
\begin{equation}\label{eq:simpleXDA}
\mathbf{X} = \mathbf{D} \mathbf{A},
\end{equation}
where
\begin{equation}
\mathbf{X} = 
    \begin{bmatrix}
    \mathbf{Y} \\
    \mathbf{Z}
    \end{bmatrix}, \quad
\mathbf{D} = 
    \begin{bmatrix}
    \mathbf{D}_{\text{c}}^{\text{xrf}} & \mathbf{D}_{\text{u}}^{\text{xrf}} & \mathbf{0} \\
    \mathbf{D}_{\text{c}}^{\text{rgb}} & \mathbf{0} & \mathbf{D}_{\text{u}}^{\text{rgb}}
    \end{bmatrix}, \quad
\mathbf{A} = 
    \begin{bmatrix}
    \mathbf{A}_{\text{c}} \\
    \mathbf{A}_{\text{u}}^{\text{xrf}} \\
    \mathbf{A}^{\text{rgb}}_{\text{u}}
    \end{bmatrix}.
\end{equation}
In this work, we begin with this more compact representation to design a model-inspired neural network architecture, as detailed in the next section.

\section{Proposed Method}

We aim to find a deep-learning architecture inspired by the mathematical formulation of the problem. Following the classic sparse representation model, we can pose the following optimization problem:
\begin{equation}\label{eq:eqL1Minim}
    \argmin_{\mathbf{A}} \lVert \mathbf{D}\mathbf{A}-\mathbf{X}\rVert^2_2 + \lambda \lVert \mathbf{A} \rVert_1, \quad \text{s.t.} \quad \mathbf{A} \geq 0,
\end{equation}
where the inequality is applied element-wise and the norms are entry-wise.

This formulation allows us to derive a neural network architecture inspired by the Iterative Shrinkage-Thresholding Algorithm (ISTA) \cite{daubechies2004iterative}. Specifically, LISTA \cite{gregor2010learning}—the learned version of ISTA—is a neural network designed such that each layer emulates one iteration of ISTA. Every layer of LISTA performs the following step from the formulation in Equation \eqref{eq:eqL1Minim}:

\begin{equation}\label{eq:listaUpdate1}
\mathbf{A}^{(k+1)} = \text{ReLU} \left( \mathbf{A}^{(k)} - \mathbf{W} \mathbf{A}^{(k)} + \mathbf{S} \mathbf{X} - \lambda \right),
\end{equation}
where $\mathbf{W}$, $\mathbf{S}$, and $\lambda$ are learnable parameters that can be optimized based on a given training dataset.

However, in our problem, the sparse representation $\mathbf{A}$ is characterized by sparsity, non-negativity, and boundedness~\cite{dai2017spatial}. To meet these requirements, we propose an unfolding network architecture incorporating a Sigmoid function with a bias term as its non-linearity. This design choice enforces the required characteristics of $\mathbf{A}$ as follows:

\begin{equation}\label{eq:listaUpdateF}
\mathbf{A}^{(k+1)} = \text{Sigmoid} \left( \mathbf{A}^{(k)} - \mathbf{W}^{(k)}\mathbf{A}^{(k)} + \mathbf{S}^{(k)}\mathbf{X} - \lambda^{(k)} \right),
\end{equation}
where $\mathbf{W}^{(k)}$, $\mathbf{S}^{(k)}$, and $\lambda^{(k)}$ are customized for each unfolded iteration $k$, thereby enhancing the capabilities of the networks without compromising its simplicity. In our model, $\lambda^{(k)}$ is not a scalar but a vector, where each component $\lambda^{(k)}_i$ corresponds to a specific row of $\mathbf{A}$. Thus, each $\lambda^{(k)}_i$ is applied element-wise to each row $i$, before the Sigmoid function. Furthermore, note that $\mathbf{Y}$, being the desired final output, is not accessible. Hence, we use $\mathbf{X} = [\mathbf{Y}_{\downarrow \uparrow}, \mathbf{Z}]^T$ as the input of the network, where $\mathbf{Y}_{\downarrow \uparrow} $ is the bilinear upscaled version of $\mathbf{Y}_{\downarrow}$.

Finally, after $K$ unfolded iterations of the network, the MA-XRF image can be obtained, leveraging Equation~\eqref{eq:simpleXDA}, as follows:
\begin{equation}\label{eq:finalHRXRF}
\hat{\mathbf{X}} = \text{Sigmoid}(\mathbf{D} \mathbf{A}^{(K)}-\lambda^{(K)} ),
\end{equation}
where $\hat{\mathbf{X}}$ is the concatenated reconstruction. We found it beneficial to include a last non-linearity with a Sigmoid layer and a bias term  $\lambda^{(K)}$ to ensure a bounded output. To extract the reconstructed HR MA-XRF image $\hat{\mathbf{Y}}$ and the HR RGB image $\hat{\mathbf{Z}}$ from $\hat{\mathbf{X}}$, slicing is performed to select specific channels:
\begin{equation}
\hat{\mathbf{Y}} = \hat{\mathbf{X}}_{[0:B]}, \quad \hat{\mathbf{Z}} = \hat{\mathbf{X}}_{[B:B+b]},
\end{equation}
where $\hat{\mathbf{X}}_{[0:B]}$ indicates slicing $\hat{\mathbf{X}}$ to select the first $B$ channels, forming the HR MA-XRF image. Similarly, $\hat{\mathbf{X}}_{[B:B+b]}$ extracts the next $b$ channels, corresponding to the HR RGB image.

\begin{figure}[]

(a)
\begin{minipage}{1.0\linewidth}
  \centering
  \centerline{\hspace*{0 cm}\includegraphics[trim=110 270 100 160, clip, width=8.5 cm]{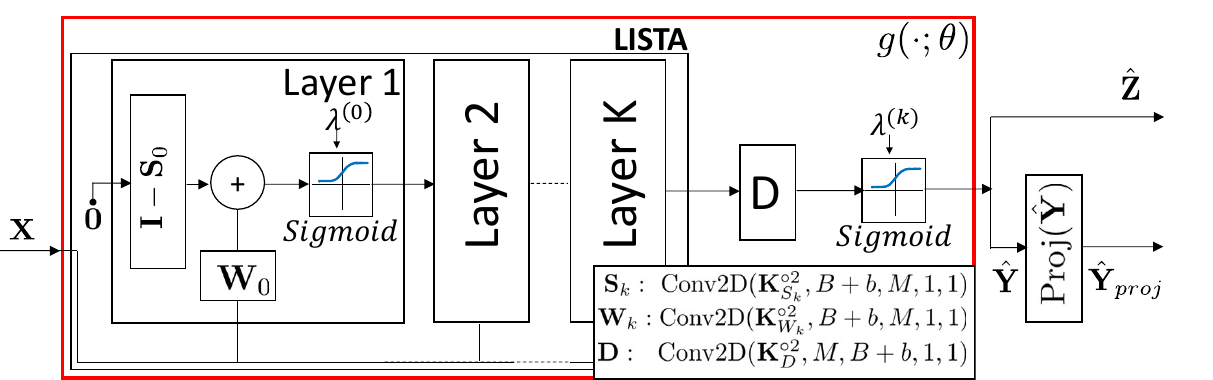}}
\end{minipage}
(b)
\begin{minipage}{1.0\linewidth}
  \centering
  \centerline{\hspace*{-0.4 cm}\includegraphics[trim=80 340 300 125, clip, height=2.9 cm]{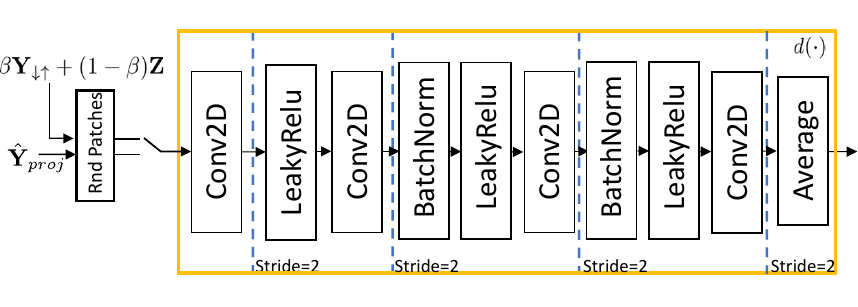}}
\end{minipage}
(c)
\begin{minipage}{1.0\linewidth}
  \centering
  \centerline{\hspace*{0 cm}\includegraphics[trim=120 315 320 150, clip, height=3.1 cm]{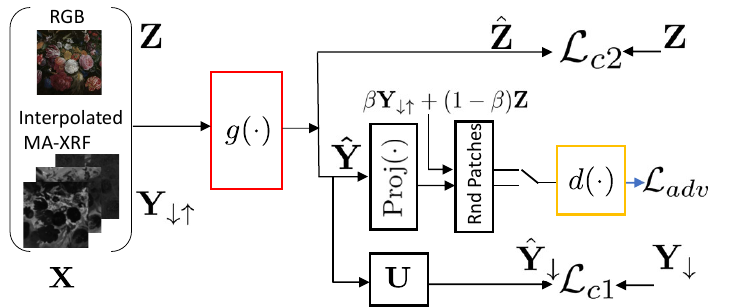}}
\end{minipage}
\caption{Overview of the method. In part (a), the architecture of the deep unfolding reconstruction network is shown, whereas part (b) presents the architecture of the discriminator. These architectures are trained within an adversarial framework, as shown in part (c).}

\label{fig:listaArch}
\end{figure}

The end-to-end network $g(\cdot;\theta)$, where $\theta$ represents the learnable parameters of the network, is shown in Figure~\ref{fig:listaArch} (a). Note that each matrix multiplication in Equation~\eqref{eq:listaUpdateF} and~\eqref{eq:finalHRXRF} is implemented with a convolutional layer with unit-length filters. The dictionary layers are required to be non-negative, as suggested in~\cite{dai2017spatial,10096322}. This constraint aids in regularizing the training process, especially given the limited availability of training data. We enforce non-negativity by squaring the weights element-wise before convolution.
\subsection{Training Strategy}

In this unsupervised approach, no separate dataset is required for training. Instead, we work directly with the input data: a HR RGB image, a LR MA-XRF image, and the downsampling matrix. Our objective is to reconstruct the corresponding HR MA-XRF image. To effectively use this information, we introduce an additional step that projects the output of the network, $\hat{\mathbf{Y}}$, onto the convex set of solutions for $\mathbf{Y}$ such that $\mathbf{Y}_{\downarrow} = \mathbf{Y}\mathbf{U}$. Specifically, we use the following projection equation \cite{meyer2023matrix}:

\begin{equation}
\text{Proj}(\hat{\mathbf{Y}}) = \hat{\mathbf{Y}} -(\hat{\mathbf{Y}}\mathbf{U} - \mathbf{Y}_{\downarrow})\mathbf{U}^T.
\end{equation}

This approach ensures that the final reconstruction always matches the downsampled version of the given LR MA-XRF image $\mathbf{Y}_{\downarrow}$. As shown in Figure~\ref{fig:listaArch} (a), the final output of our reconstruction method is $\text{Proj}(\hat{\mathbf{Y}})$. Furthermore, to effectively guide the training, we employ a regularized loss function that integrates both fidelity to the observed data and regularization, as follows:

\begin{equation}\label{eq:trainloss}
\mathcal{L}_{c1}(\mathbf{Y}_{\downarrow},\mathbf{\hat{Y}}_{\downarrow}) + \alpha_1 \mathcal{L}_{c2}(\mathbf{Z},\mathbf{\hat{Z}}) + \alpha_2 \mathcal{L}_{adv}(\text{Proj}(g(\mathbf{X}))),
\end{equation}

where $\mathbf{Y}_{\downarrow}$ is the given LR image, $\mathbf{\hat{Y}}_{\downarrow}$ is the downsampled network reconstruction (without the projection step), $\mathbf{Z}$ is the given HR RGB image, and $\mathbf{\hat{Z}}$ is the reconstructed HR RGB image. Here, $\mathbf{X}$ represents the given concatenated input, as in Equation~\eqref{eq:listaUpdateF}. The weight for each loss component is controlled by the scalars $\alpha_1$ and $\alpha_2$. $\mathcal{L}_{adv}(\text{Proj}(g(\mathbf{X})))$ represents an adversarial regularizer~\cite{verinaz2023physics,lunz2018adversarial}, enforcing the network towards generating more realistic HR MA-XRF images. The adversarial loss $\mathcal{L}_{adv}(\cdot)$ is calculated using the discriminator $d(\cdot)$ from Figure~\ref{fig:listaArch} (b) based on least squares adversarial training~\cite{mao2017least}. We adopt Mean Squared Error (MSE) for both $\mathcal{L}_{c1}(\cdot)$ and $\mathcal{L}_{c2}(\cdot)$. 

During adversarial training, we specifically target patches misclassified by the model to exclusively focus on errors that contribute to model improvement. Misclassification is determined by the output of the discriminator, $d(\mathbf{x}_{\text{patch}})$, and a reference value $\tau$. A patch $\mathbf{x}_{\text{patch}}$ is considered misclassified if $d(\mathbf{x}_{\text{patch}}) < \tau$ when $\tau > 0$, or $d(\mathbf{x}_{\text{patch}}) > \tau$ when $\tau < 0$. For each specified $\tau$, which is set to 1 or -1 in our work, the distribution of misclassified real MA-XRF patches is denoted by $\mathbb{P}_{r}^\tau$, and the distribution of misclassified reconstructed patches by $\mathbb{P}_{\theta}^\tau$. Then, the adversarial loss is formulated as:
\begin{equation}\label{eq:advLoss}
\mathcal{L}_{adv}(\mathbf{x}_{\text{patch}}) = \mathbb{E}_{\mathbf{x}_{\text{patch}} \sim \mathbb{P}_{\theta}^1} \left[ \left(d(\mathbf{x}_{\text{patch}}) - 1 \right)^2 \right].
\end{equation}
In practice, this loss is implemented by first sampling a batch of patches and then removing those that are correctly classified before computing the mean. The training of the discriminator also employs this focused approach:
\begin{equation}\label{eq:warsssLoss}
\mathbb{E}_{\mathbf{x}_{\text{patch}} \sim \mathbb{P}_{\theta}^{-1}} \left[ \left(d(\mathbf{x}_{\text{patch}}) + 1 \right)^2 \right] + \mathbb{E}_{\mathbf{x}_{\text{patch}} \sim \mathbb{P}_r^1} \left[ \left(d(\mathbf{x}_{\text{patch}}) - 1 \right)^2 \right],
\end{equation}
We have empirically found that this formulation improves stability and training efficiency by focusing the loss calculation on truly misclassified patches.

We utilize the spatial similarities present in RGB images to avoid the need for real patches for training. Specifically, we create pseudo-real patches by computing a weighted average between a randomly selected channel of the interpolated MA-XRF image and the channel of the RGB image that is most correlated with it, as follows:
\begin{equation}\label{eq:realImagesInit}
\beta \mathbf{Y_{\downarrow\uparrow\text{patch}}} + (1-\beta) \mathbf{Z}_{_\text{patch}},
\end{equation}
where $\mathbf{Y_{\downarrow\uparrow\text{patch}}}$ is a patch of the upscaled version of the HR MA-XRF image, and $\mathbf{Z}_{\text{patch}}$ is a patch from the RGB image channel that shows the highest correlation with the MA-XRF channel. See Figure~\ref{fig:listaArch}~(c).
	
\section{Experiments and Results}
\label{sec:results}
\subsection{Implementation details}
To evaluate the effectiveness of our method, we conducted tests on datasets derived from three renowned oil paintings. These include Jan Davidsz. de Heem's \textit{Flowers and Insects} (oil on canvas, Royal Museum of Fine Arts Antwerp, inv. no. 54) \cite{de2017jan};  Francisco de Goya's \textit{Doña Isabel de Porcel} (before 1805, oil on canvas, The National Gallery, London, NG1473) \cite{NG_Technical_37}; and Leonardo da Vinci's \textit{The Virgin of the Rocks} (circa 1491/2-1508, oil on poplar, thinned and cradled, The National Gallery, London, NG1093) \cite{NG_Technical_41}. Our approach was benchmarked against several methods including SSR \cite{dai2017spatial} and SSRCU~\cite{10096322} specifically designed for MA-XRF super-resolution; CSTF \cite{li2018fusing}, CMS \cite{zhang2018exploiting}, LTTR \cite{dian2019learning} targeting HSI super-resolution; and CAR \cite{sun2020learned}, HAT \cite{chen2022activating}, Swin2SR \cite{conde2022swin2sr} for single image super-resolution (SISR). We used deconvoluted MA-XRF element maps~\cite{yan2023fast} as HR ground truth, with LR versions produced by 4$\times$ downsampling.

Next, we detail the specific parameters and configurations used in our experiments. The MA-XRF image has $B=21$, $B=7$, and $B=9$ spectral channels for \textit{Flowers and Insects}, \textit{Doña Isabel de Porcel}, and \textit{The Virgin of the Rocks}, respectively. All RGB images are $512\times 512\times 3$ ($b=3$). The number of channels $M$ in each layer of our DNN is set to $64$, and the number of layers $K$ is set to $5$. See Figure~\ref{fig:listaArch}~(a). 

The initial training phase excludes the adversarial loss component specified in Equation~\eqref{eq:trainloss}. During this pretraining phase, spanning $1\times10^5$ epochs with Adam optimizer, we set $\alpha_1$ at $0.003$. This phase includes augmented data from the same given inputs, $\mathbf{Y}{\downarrow\downarrow}$, a twice downsampled HR MA-XRF image, and $\mathbf{Z}{\downarrow}$, the corresponding RGB image. To address vanishing gradients, this stage excludes the final Sigmoid layer, and weights are constrained to remain below one for enhanced generalization.

In the adversarial training stage, we maintain the weight parameter $\alpha_1$ and introduce $\alpha_2$, set at $0.5\times10^{-6}$. The adversarial training uses a patch size and batch size of 32, with learning rates of $3\times10^{-4}$ for the generator $g(\cdot)$ and $3\times10^{-6}$ for the discriminator $d(\cdot)$, spanning $2\times10^6$ epochs. The scalar $\beta$ in Equation~\eqref{eq:realImagesInit} is randomly adjusted from 0 to 0.9 in each iteration, aiding data augmentation. We also implement random flipping of patches to further diversify the adversarial training. During this phase, the network architecture and data are as described in previous section. Finally, the discriminator architecture follows the standard design shown in Figure~\ref{fig:listaArch}~(b).

\subsection{Results}

\begin{table*}[t]
\centering
\caption{Comparative results for 4$\times$ upscaling on datasets from three Old Master paintings, presented in terms of RMSE and PSNR for various super-resolution methods. Best results are highlighted in bold and second-best are underlined.}

\begin{tabular}{c | c | c c c | c c c | c c c }
    \hline
    \multirow{3}{*}{\shortstack{Dataset}} & \multirow{3}{*}{\shortstack{Metric}} & \multicolumn{8}{c}{\shortstack{Methods for SR problems}} \\
    \cline{3-11}
    & & \multicolumn{3}{c|}{SISR} & \multicolumn{3}{c|}{HSI SR} & \multicolumn{3}{c}{MA-XRF SR} \\
    \cline{3-11}
     & & CAR \cite{sun2020learned} & HAT \cite{chen2022activating} & Swin2SR \cite{conde2022swin2sr} & CSTF \cite{li2018fusing} & CMS \cite{zhang2018exploiting} & LTTR \cite{dian2019learning} & SSR \cite{dai2017spatial} & SSRCU~\cite{10096322} & Ours \\
    \hline
    \multirow{2}{*}{\shortstack{\textit{Flowers and} \\ \textit{Insects}}} & RMSE & 0.0380 & 0.0275 & 0.0275 & 0.1336 & 0.0412 & 0.0582 & 0.0232 & \underline{0.0187} & \textbf{0.0145} \\
    & PSNR & 28.42 & 31.22 & 31.22 & 17.48 & 27.70 & 24.71 & 32.69 & \underline{34.55} & \textbf{36.75} \\
    \hline
    \multirow{2}{*}{\shortstack{\textit{The Virgin of} \\ \textit{the Rocks}}} & RMSE & 0.0281 & 0.0226 & 0.0229 & 0.0771 & 0.0397 & 0.0657 & 0.0223 & \underline{0.0182} & \textbf{0.0163} \\
    & PSNR & 31.01 & 32.92 & 32.79 & 22.26 & 28.03 & 23.65 & 33.03 & \underline{34.80} & \textbf{35.77} \\
    \hline
    \multirow{2}{*}{\shortstack{\textit{Doña Isabel} \\ \textit{de Porcel}}} & RMSE & 0.0388 & 0.0296 & 0.0297 & 0.0777 & 0.0373 & 0.0513 & 0.0264 & \underline{0.0252} & \textbf{0.0210} \\
    & PSNR & 28.22 & 30.57 & 30.54 & 22.19 & 28.56 & 25.80 & 31.55 & \underline{31.98} & \textbf{33.57} \\
    \hline
\end{tabular}
\label{table:RMSE_PSNR}
\end{table*}

\begin{figure*}[!t]
\centering
\begin{minipage}{0.2\linewidth}
\centering
    \begin{subfigure}{1\linewidth}
        \centering
        \includegraphics[width=1\linewidth]{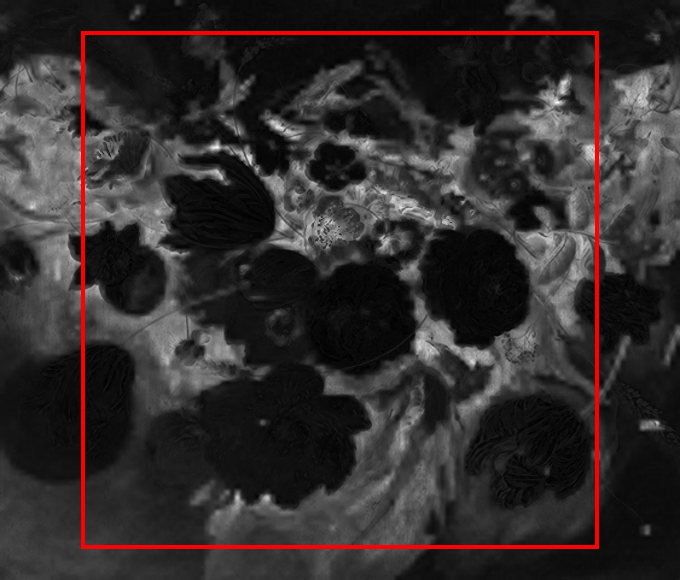}
        \caption{GT HR Ca $K_{\alpha}$ map}
    \end{subfigure}
    \\
    \begin{subfigure}{1\linewidth}
        \centering
        \includegraphics[width=1\linewidth]{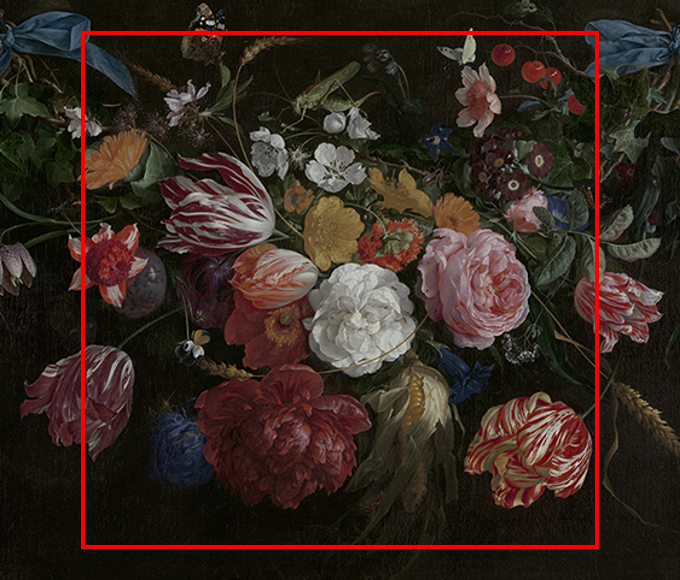}
        \caption{HR RGB}
    \end{subfigure}
\end{minipage}
\begin{minipage}{0.74\linewidth}
\centering
    \begin{subfigure}{0.19\linewidth}
        \centering
        \includegraphics[width=\linewidth]{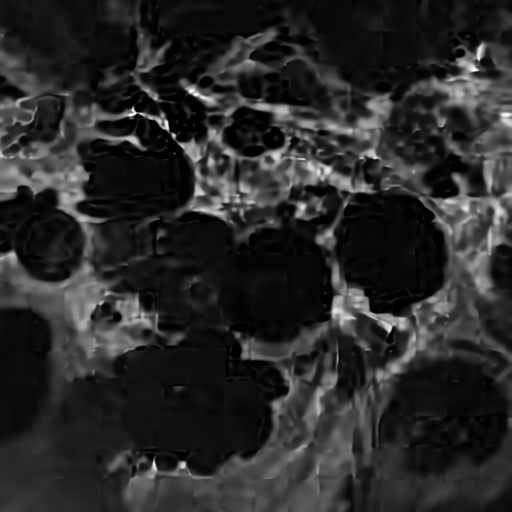}
        \includegraphics[width=\linewidth]{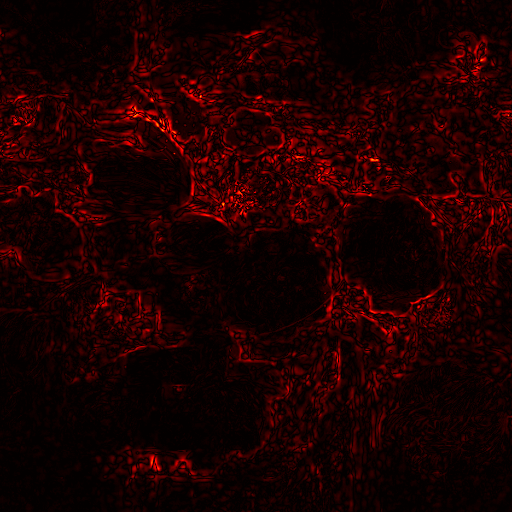}
        \caption{CAR \cite{sun2020learned}}
    \end{subfigure}
    \begin{subfigure}{0.19\linewidth}
        \centering
        \includegraphics[width=\linewidth]{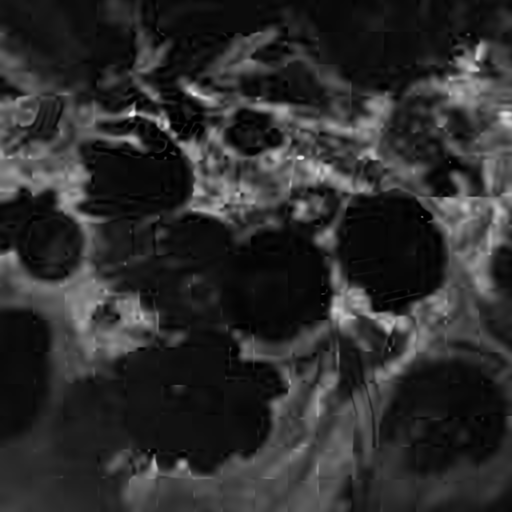}
        \includegraphics[width=\linewidth]{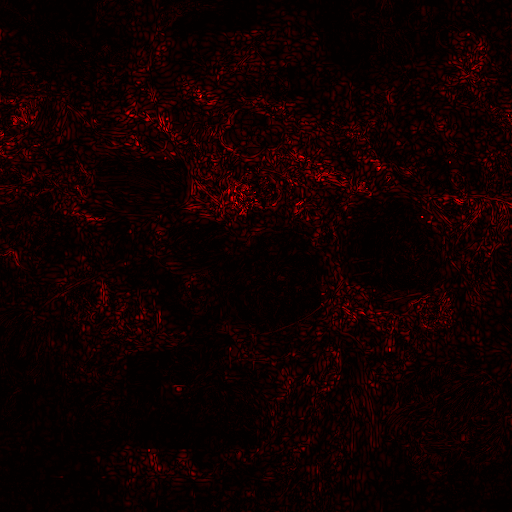}
        \caption{HAT \cite{chen2022activating}}
    \end{subfigure}
    \begin{subfigure}{0.19\linewidth}
        \centering
        \includegraphics[width=\linewidth]{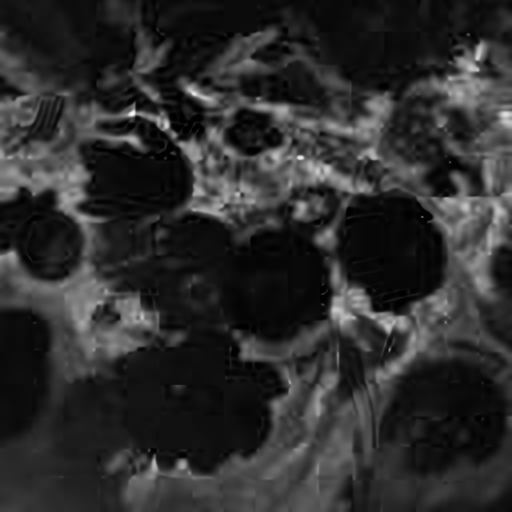}
        \includegraphics[width=\linewidth]{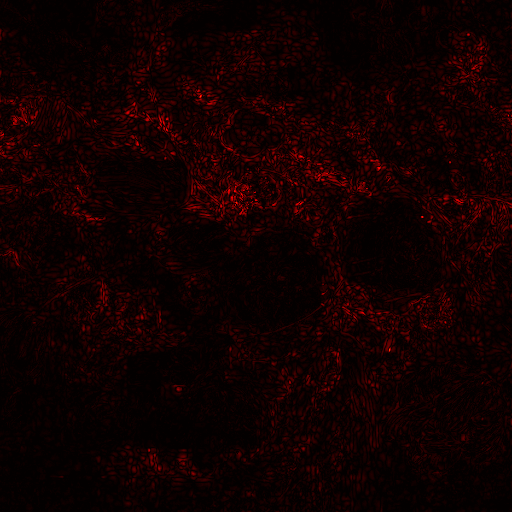}
        \caption{Swin2SR \cite{conde2022swin2sr}}
    \end{subfigure}
    \begin{subfigure}{0.19\linewidth}
        \centering
        \includegraphics[width=\linewidth]{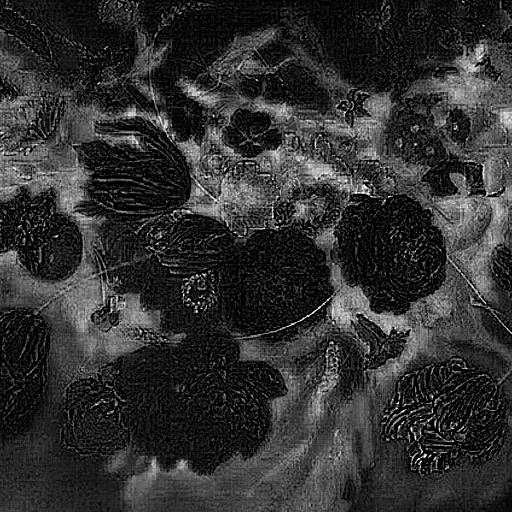}
        \includegraphics[width=\linewidth]{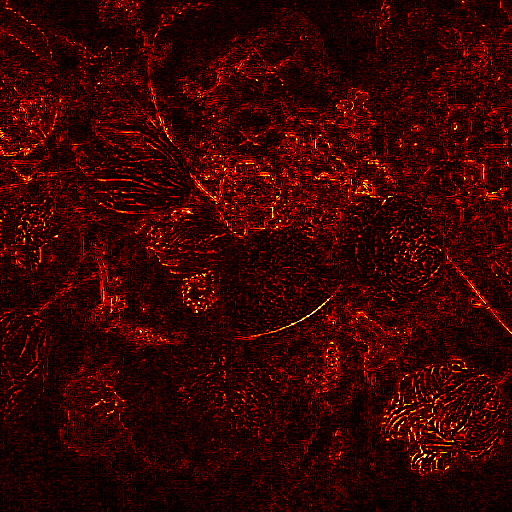}
        \caption{CSTF \cite{li2018fusing}}
    \end{subfigure}
    \\
    \begin{subfigure}{0.19\linewidth}
        \centering
        \includegraphics[width=\linewidth]{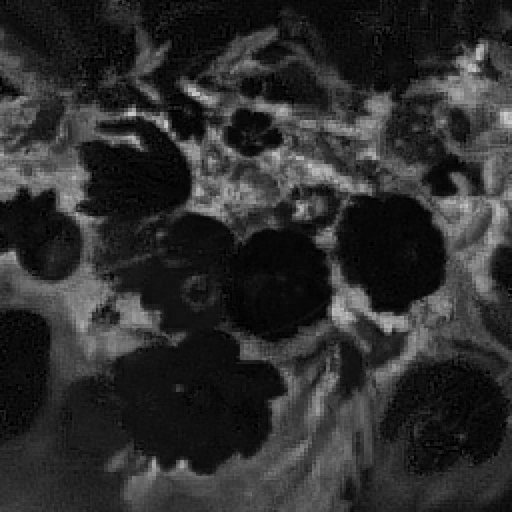}
        \includegraphics[width=\linewidth]{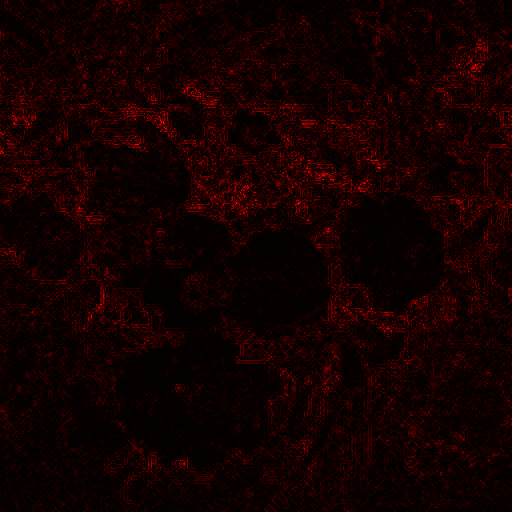}
        \caption{CMS \cite{zhang2018exploiting}}
    \end{subfigure}
    \begin{subfigure}{0.19\linewidth}
        \centering
        \includegraphics[width=\linewidth]{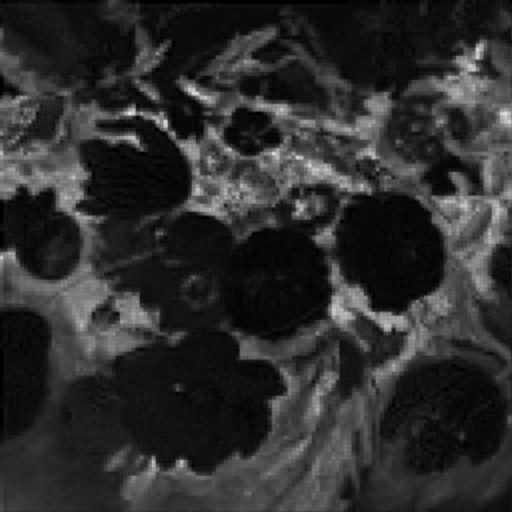}
        \includegraphics[width=\linewidth]{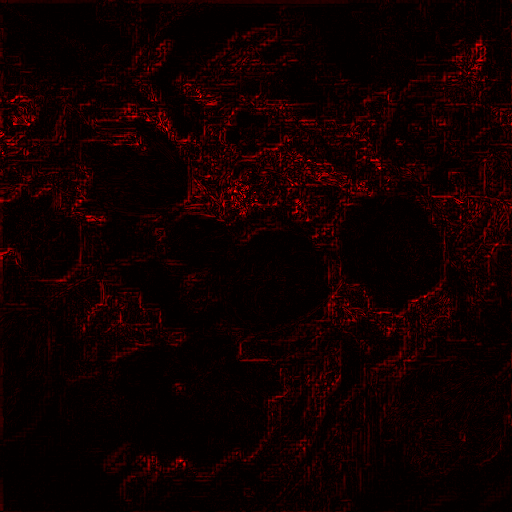}
        \caption{LTTR \cite{dian2019learning}}
    \end{subfigure}
    \begin{subfigure}{0.19\linewidth}
        \centering
        \includegraphics[width=\linewidth]{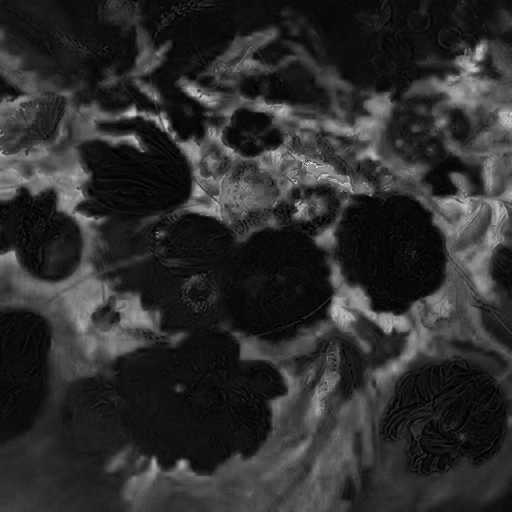}
        \includegraphics[width=\linewidth]{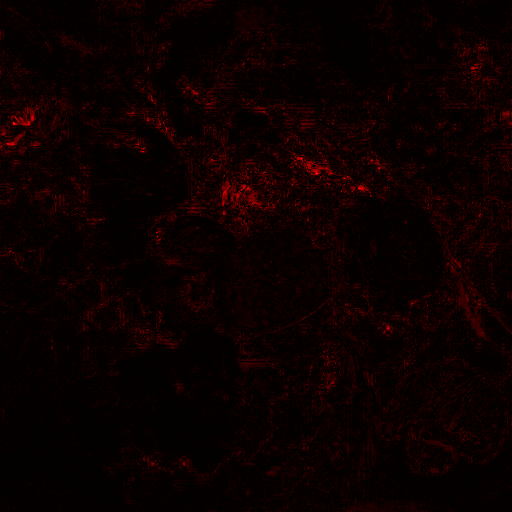}
        \caption{SSR \cite{dai2017spatial}}
    \end{subfigure}
    \begin{subfigure}{0.19\linewidth}
        \centering
        \includegraphics[width=\linewidth]{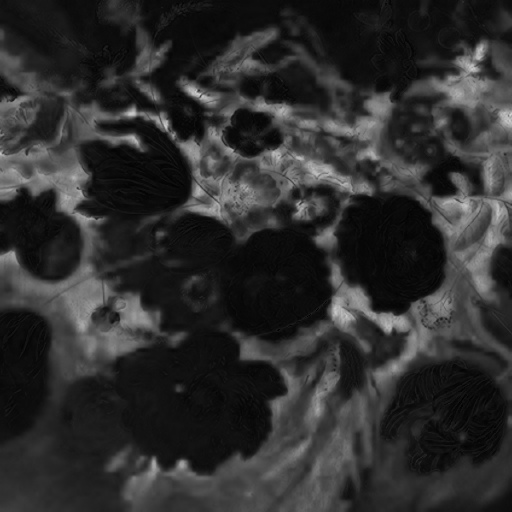}
        \includegraphics[width=\linewidth]{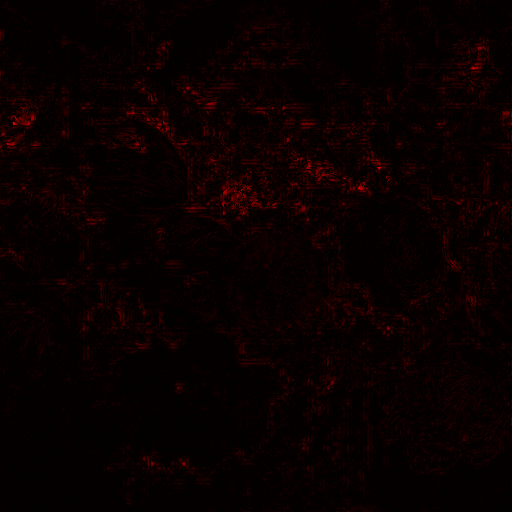}
        \caption{SSRCU~\cite{10096322}}
    \end{subfigure}
    \begin{subfigure}{0.19\linewidth}
        \centering
        \includegraphics[width=\linewidth]{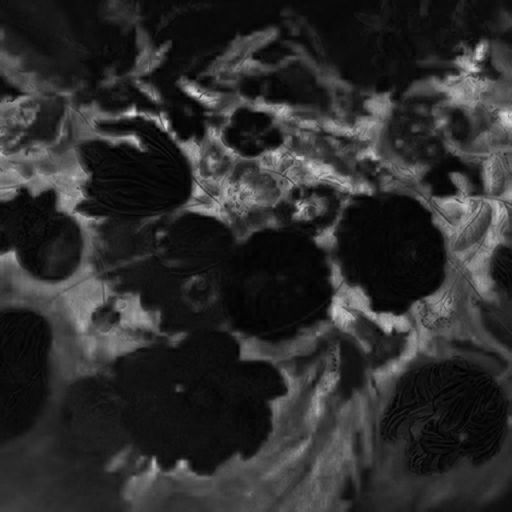}
        \includegraphics[width=\linewidth]{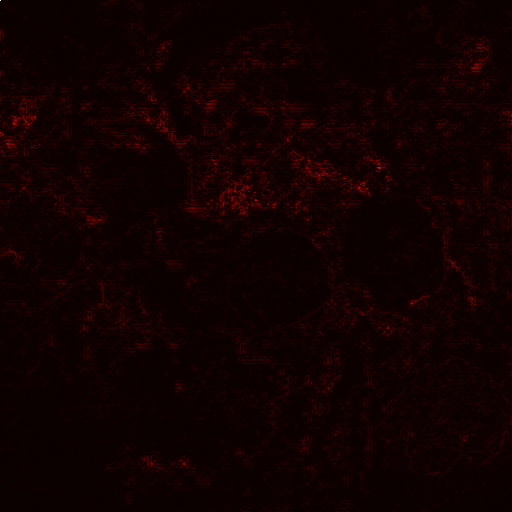}
        \caption{Ours}
    \end{subfigure}
\end{minipage}
\caption{Visual comparison of HR MA-XRF data (Ca $K_{\alpha}$ element distribution maps), reconstructed using various super-resolution methods for a 4$\times$ upscaling. Part (a) and (b) show the ground truth HR Ca $K_{\alpha}$ map and the HR RGB image of \textit{Flowers and Insects} $\copyright$KMSKA, respectively. Part (c) to (k) illustrate the reconstructed maps and their corresponding error maps for each method.}
\label{fig:Ca_maps}
\end{figure*}

The proposed approach shows superior performance across all evaluated datasets, notably improving upon SSRCU~\cite{10096322}, a state-of-the-art MA-XRF super-resolution technique. Table \ref{table:RMSE_PSNR} presents the evaluation results for various super-resolution methods, comparing their performance in terms of Root Mean Squared Error (RMSE) and Peak Signal-to-Noise Ratio (PSNR). Among them, the SISR method HAT~\cite{chen2022activating} stands out in the Single Image Super-Resolution (SISR) and Hyperspectral Image Super-Resolution (HSI SR) categories, but these methods still fall short compared to other MA-XRF-specific super-resolution techniques. Note that although HSI SR methods are designed for multimodal super-resolution, they do not perform better than SISR in this experiment. Differences in spectral ranges, as well as the physical properties of the imaging techniques, such as XRF detecting only certain elements, highlight the need for methods specifically tailored for MA-XRF super-resolution. In particular, our approach demonstrates the best performance in all the metrics.

Further insights into the performance improvements are visible in Fig \ref{fig:Ca_maps}, which displays the calcium (Ca $K_{\alpha}$) element distribution maps and their corresponding error maps. Our method enables the recovery of fine details and sharper edges through a model-based network architecture and a specialized adversarial loss, resulting in higher-quality MA-XRF reconstructions.

\section{Conclusion}
\label{sec:conclusion}
This paper introduces the first deep learning approach tailored to MA-XRF super-resolution, specifically designed for the analysis of Old Master paintings. Our approach requires only a single HR RGB image alongside LR MA-XRF data for training, thus removing dependency on extensive datasets or pre-trained architectures. Both qualitative assessments of image quality and quantitative evaluations using metrics such as PSNR and RMSE confirm that our method surpasses various state-of-the-art techniques. The efficiency of our method, outlined by minimal data requirements, can enable broader application in the analysis of Old Master paintings. This includes techniques such as Macro X-ray Powder Diffraction and Macro Fourier Transform Infrared Scanning in reflection mode.

\section{Acknowledgement}

The authors would like to thank the Royal Museum of Fine Arts Antwerp (KMSKA) and Prof. Koen Janssens and Nouchka De Keyser from University of Antwerp for providing the \textit{Flowers and Insects} data.

\bibliographystyle{IEEEbib.bst}
{\bibliography{reference.bib}}

\end{document}